\documentclass[aps,amsmath,amssymb,twocolumn,10pt,prl,superscriptaddress]{revtex4-1}
\usepackage{graphicx}  
\usepackage[dvipsnames]{xcolor}
\usepackage[colorlinks=true,linkcolor=blue,citecolor=red,urlcolor=magenta]{hyperref}
\usepackage{charter} 

\begin{document}
\title{Universality of internal correlations \\ of strongly interacting $\mathfrak{p}$-wave fermions in one-dimensional geometry}
\author{Przemys{\l}aw Ko\'scik}
\affiliation{Department of Computer Sciences, University of Applied Sciences, ul. Mickiewicza 8, PL-33100 Tarn\'{o}w, Poland}
\author{Tomasz Sowi\'nski} 
\affiliation{Institute of Physics, Polish Academy of Sciences, Aleja Lotnik\'ow 32/46, PL-02668 Warsaw, Poland} 
\date{\today} 

\begin{abstract}
We consider the many-body ground state of polarized fermions interacting via zero-range $\mathfrak{p}$-wave forces in a one-dimensional geometry. We rigorously prove that in the limit of infinite attractions spectral properties of any-order reduced density matrix describing arbitrary subsystem are completely independent of the shape of an external potential. It means that quantum correlations between any two subsystems are in this limit insensitive to the confinement. In addition, we show that the purity of these matrices quantifying the amount of quantum correlations can be obtained analytically for any number of particles without diagonalizing them. This observation may serve as a rigorous benchmark for other models and methods describing strongly interacting $\mathfrak{p}$-wave fermions.
\end{abstract}
\maketitle

{\bf Motivation.} Over the last two decades, there has been growing interest in the properties of one-dimensional systems composed of particles confined in trapping potentials, which are described by generic Hamiltonians of the form \begin{equation}\label{Hamiltonian_total}
 {\cal H} = \sum_{i=1}^N\left(-\frac{1}{2}\frac{\partial^2}{\partial x_i^2} + V(x_i) + \sum_{j=i+1}^NU(x_i-x_j)\right),
\end{equation}
where $V(x)$ and $U(x)$ are the trapping and interacting potentials, respectively. In particular, in the light of experimental progress with ultracold atomic systems, much effort has been devoted to a better understanding of properties of systems of bosons and two-component fermionic mixtures interacting via $\mathfrak{s}$-wave zero-range interactions, $U(x)=g\delta(x)$ (for general reviews see \cite{2012BlumeRPP,2013GuanRevModPhys,2016ZinnerRev,2019SowinskiRPP,2022MistakidisReview}). In contrast, despite the growing experimental activity~\cite{2003RegalPRL,2004ZhangPRA,2005GunterPRL,2013NakasujiPRA,2021TopPRA,2023VenuNature}, relatively little theoretical attention has been paid to one-component systems of polarized fermions interacting via zero-range $\mathfrak{p}$-wave forces~\cite{1986SebaRMP,2003SenJPhysA,2004GirardeauPRA,2004KanjilalPRA,2006GirardeauPRL,2006delCampoPRA,2007HaoPRA}. In one-dimensional geometry, they are represented by the following differential operator
\begin{equation}
U(x)=-\frac{g_F}{2}\overleftarrow{\frac{\partial}{\partial x}}\delta(x)\overrightarrow{\frac{\partial}{\partial x}},
\end{equation}
 where $g_F$ is the effective $\mathfrak{p}$-wave interaction strength. One of the fundamental results on $\mathfrak{p}$-wave fermions in one dimension, following observation by Girardeau \cite{1960GirardeauJMP}, was given in \cite{1999CheonPRL}. It was rigorously proven that, independently of the trapping potential, for any $g_F$ the many-body ground-state $\Psi(x_1,\ldots,x_N)$ of the Hamiltonian \eqref{Hamiltonian_total} can be derived directly from the many-body ground-state $\Psi_B(x_1,\ldots,x_N)$ of one-component $\mathfrak{s}$-wave bosonic system obtained for interaction strength $g=-2/g_F$ via anti-symmetrization transformation of the form
$
\Psi(x_{1},..,x_{N})={\cal A}(x_{1},..,x_{N})\Psi_{B} (x_{1},..,x_{N}),
$
where ${\cal A}(x_{1},..,x_{N})=\Pi_{i<j}\mathrm{sgn}(x_{i}-x_{j})$. This mapping turned out to be exceptionally useful for systems confined in a homogenous box potential where Bethe ansatz can be exploited~\cite{2006HaoPRA,2007HaoPRA}. Although, in principle, the transformation gives a route to determine the ground-state wave function of $\mathfrak{p}$-wave fermions from the corresponding bosonic system, due to its non-trivial structure there is no direct mapping of different properties between these systems. Particularly, mutual correlations encoded in reduced density matrices cannot be deduced from correlations in bosonic counterparts.

The mapping is particularly convenient in limiting interaction strengths. For example, the wave function of bosonic gas in the Tonks-Girardeu limit ($g\rightarrow +\infty$) can be easily obtained from the many-body wave function of non-interacting fermions expressed as symmetrized Slater determinant of $N$ single-particle orbitals determined by potential $V(x)$. This observation triggered a progress in better understanding of strongly repulsive bosonic systems. On the opposite, the ground-state of infinitely strongly attracting $\mathfrak{p}$-wave fermions ($g_F\rightarrow -\infty$) is mapped from the non-interacting ground-state of bosonic system, {\it i.e.}, it depends only on one function $\phi(x)$ being the lowest eigenstate of a single-particle Hamiltonian. In this limit the many-body ground-state wave function of $\mathfrak{p}$-wave fermions has a form
\begin{equation} \label{wf}
\Psi(x_{1},..,x_{N})= \prod_{i=1}^N\left(\phi(x_{i})\prod_{j=i+1}^N\mathrm{sgn}(x_{i}-x_{j})\right).
\end{equation}
It suggests that knowledge of the ground-state function $\phi(x)$ is necessary to determine any properties of $\mathfrak{p}$-wave fermions in the limit of strong attraction $g_F\rightarrow -\infty$. Thus, to get quantitative predictions, typically one assumes that external potential is uniform~\cite{2005BenderPRL} or parabolic~\cite{2006MinguzziPRA,2020KoscikNJP}. We show in the following that {\it this kind of assumption is not needed if any internal correlations} between subsystems are considered. 

{\bf Internal correlations.} The most general object encoding internal correlations in any many-body system of indistinguishable particles is the whole set of $p$-particle reduced density matrices ($p$-RDM)~\cite{2002EckertAnnPhys,2012MazziottiPRL,2022KnightNJP}. They are obtained by considering the subsystem of $p$ particles and by integrating out remaining part from the density matrix
 \begin{align}
 \rho^{(p)}(\boldsymbol{x}_{p},\boldsymbol{x}'_{p})&=\int \mathrm{d}\boldsymbol{q}_p\Psi^{*}(\boldsymbol{x}_{p},\boldsymbol{q}_{p})\Psi(\boldsymbol{x}'_{p},\boldsymbol{q}_p),
 \end{align}
 where $\boldsymbol{x}_{p}=(x_1,\ldots,x_{p})$ and $\boldsymbol{q}_{p}=(x_{p+1},\ldots,x_N)$. Utilizing the known form of the many-body ground state wave function \eqref{wf} it is straightforward to show that $p$-RDM can be written as
\begin{multline} \label{exactrho}
\rho^{(p)}(\boldsymbol{x}_{p},\boldsymbol{x}'_{p})= \mathrm{P}(\boldsymbol{x}_{p},\boldsymbol{x}'_{p})^{N-p} \\ \times\prod_{i=1}^{p}\phi(x_{i})\phi(x'_{i})
 \prod_{j=i+1}^{p}\mathrm{sgn}(x_{i}-x_{j})\mathrm{sgn}(x'_{i}-x'_{j}),
 \end{multline}
where
\begin{align}
\mathrm{P}(\boldsymbol{x}_{p},\boldsymbol{x}'_{p})&=\int\mathrm{d}z\,\phi^2(z)\prod_{i=1}^{p}\mathrm{sgn}(x_{i}-z)\mathrm{sgn}(x'_{i}-z).
\end{align}
From the physical perspective, it is extremely useful to represent any $p$-RDM in a diagonal form as
\begin{equation}\label{diag}
\rho^{(p)}(\boldsymbol{x}_{p},\boldsymbol{x}'_{p})=\sum_k \lambda^{(p)}_{k} u_{k}(\boldsymbol{x}_{p})u^{*}_{k}(\boldsymbol{x}'_{p}),
\end{equation}
where eigenvalues $\lambda^{(p)}_k$ and eigenorbitals $u_k(\boldsymbol{x}_p)$ fulfill the integral eigenequation 
\begin{equation}\label{calg}
\int\mathrm{d}\boldsymbol{x}_p'\,\rho^{(p)}(\boldsymbol{x}_p,\boldsymbol{x}'_p)u_{k}(\boldsymbol{x}_p')=\lambda^{(p)}_{k} u_{k}(\boldsymbol{x}_p).
\end{equation}
Eigenvalues $\lambda^{(p)}_{k}$ directly determine quantum entanglement between parties after dividing the system into $p$ and $N-p$ particles. They are also used to identify different structures of the many-body states, such as $p$-order coherence and   fragmentation~\cite{2008SakmannPRA,2016LodePRA} or off-diagonal long-range order to indicate the collective formation of pairs ($p=2$)~\cite{1962YangRMP} or triples ($p=3$)~\cite{2022GuoPRA,2023GuoPRB}. It is also clear that if the eigenvalues are known for any $p$, complete knowledge of internal correlations in the system can be extracted. There are a plethora of different quantities describing correlations. In the following, we use one of the simplest -- the participation number defined as $K^{(p)}=1/\mathrm{Tr}\left[(\rho^{(p)})^2\right]$~\cite{1994GrobeJPhysB}. This number, being the inverse of the quantum purity ${\cal P}^{(p)}=\mathrm{Tr}\left[(\rho^{(p)})^2\right]=\sum_k \left(\lambda^{(p)}_k\right)^2$, can be viewed as an effective number of eigenorbitals decomposing the reduced density matrix $\rho^{(p)}$. It is worth mentioning that, although these quantities are introduced purely on theoretical grounds, there are experimental paths to capture them~\cite{2015IslamNature,2016HaukeNatPhys}.

{\bf Universality of $\mathfrak{p}$-wave fermions.} It is clear that the ground function $\phi(x)$ enters to the expression of any $p$-RDM in a very non-trivial way. Therefore one suspects that their spectral properties are strongly related to the external potential $V(x)$. This reasoning is also very intuitive since in other one-dimensional systems we observe a strong influence of external confinement on internal entanglement. Particularly, it is true for the Tonks-Girardeu gas of infinitely repulsively interacting bosons. In contrast to this intuitive picture in the following we rigorously show that eigenvalues $\lambda_k^{(p)}$ are completely independent of the confinement for arbitrary $p$, while corresponding eigenorbitals $u_k(\boldsymbol{x}_p)$ obtained for different confinements are related via a straightforward, analytical transformation. 

To present reasoning as clearly as possible, let us first show the universality of the decomposition for the $1$-RDM. In this case, expression \eqref{exactrho} simplifies to the known form~\cite{2005BenderPRL}
\begin{multline}\label{1rdm}
\rho^{(1)}(x,x')= \phi(x)\phi(x') \\
\times\left(\int\,\mathrm{d}z\,\phi^2(z)\,\mathrm{sgn}(x-z)\mathrm{sgn}(x'-z)\right)^{N-1}.
\end{multline}
Now, let us consider the cumulative distribution function $F(x)$ built from the single-particle distribution $\phi^2(x)$ defined as
\begin{equation} \label{Fcumul}
F(x) = \int_{-\infty}^x \mathrm{d}z\, \phi^2(z).
\end{equation}
By performing a simple change of variables $\xi = F(z)$ ($\mathrm{d}\xi=\phi^2(z)\mathrm{d}z$), $y=F(x)$, and $y'=F(x')$  and making a transformation $u_{k}(x)=\phi(x)v_{k}(F(x))$  one finds that the original eigenproblem
\begin{equation} \label{1rdmorigin}
\int\mathrm{d}x'\,\rho^{(1)}(x,x')u_{k}(x')=\lambda^{(1)}_{k} u_{k}(x)
\end{equation}
is uniquely transformed to another $\phi$-independent eigenproblem 
\begin{equation} \label{newep}
\int_{0}^{1}\mathrm{d}y'\,\rho^{(1)}_{0}(y,y')v_{k}(y')=\lambda^{(1)}_{k} v_{k}(y),
\end{equation}
where
\begin{multline} \nonumber
\rho^{(1)}_0(y,y')= \\
\left[\int_{0}^{1}\!\!\mathrm{d}\xi\,\mathrm{sgn}\!\left(F^{-1}(y)\!-\!F^{-1}(\xi)\right)\mathrm{sgn}\!\left(F^{-1}(y')\!-\!F^{-1}(\xi)\right)\right]^{N-1}.
\end{multline}
Now, we can utilize the fact that the distribution \eqref{Fcumul} and its inverse are growing functions of their arguments  ($\mathrm{sgn}(F^{-1}(a)-F^{-1}(b))=\mathrm{sgn}(a-b)$). This simplifies expression further to manifestly $\phi$-independent form
\begin{align}
\rho^{(1)}_0(y,y')&=\left (\int_{0}^{1}\!\!d\xi\, \mathrm{sgn}(y-\xi)\,\mathrm{sgn}(y' - \xi)\right)^{N-1} \nonumber \\ &=\left(1-2\left|y-y'\right|\right)^{N-1}.
\end{align}
It is quite easy to show that the proposed transformation preserves the inner product between orbitals,
\begin{multline}\label{wa}
\int\mathrm{d}x\, u_{k}(x)u_{k'}(x) \\
=\int\mathrm{d}x\, v_{k}(F(x))v_{k'}(F(x))\phi^2(x) = \\
\int_{0}^{1}\mathrm{d}y\, v_{k}(y)v_{k'}(y).
\end{multline}
Importantly, the original density matrix $\rho^{(1)}(x,x')$ has exactly the same eigenvalues $\lambda^{(1)}_k$ as the transformed density matrix $\rho_0^{(1)}(y,y')$. All that means that the task of solving the original eigenproblem \eqref{1rdmorigin} can be always reduced to the simpler task of solving the universal $\phi$-independent eigenproblem \eqref{newep}. Consequently, the spectral properties of the $1$-RDM \eqref{1rdm} {\it do not} depend on the ground function $\phi(x)$, {\it ergo} they do not depend on the shape of the external potential $V(x)$. 

Essentially, the proof for higher $p$-RDMs is very analogous. The transformation of variables from $\boldsymbol{x}_p$ to $\boldsymbol{y}_p$ should be performed on all vector elements simultaneously, {\it i.e.}, $\boldsymbol{y}_p=F(\boldsymbol{x}_p)$ and $\boldsymbol{y}'_p=F(\boldsymbol{x}'_p)$. After that (utilizing again monotonicity of the cumulative distribution) the original eigenproblem \eqref{calg} is transformed into the form
\begin{equation} \label{eigenprob2}
\int_{\boldsymbol{0}}^{\boldsymbol{1}}\mathrm{d}\boldsymbol{y}_p'\,\rho^{(p)}_0(\boldsymbol{y}_p,\boldsymbol{y}'_p)v_{k}(\boldsymbol{y}_p')=\lambda^{(p)}_{k} v_{k}(\boldsymbol{y}_p).
\end{equation}
with the transformed $p$-RDM being independent on the ground function $\phi(x)$
\begin{multline} \label{rhopp}
\rho_0^{(p)}(\boldsymbol{y}_{p},\boldsymbol{y}'_{p})= \mathrm{P}_0(\boldsymbol{y}_{p},\boldsymbol{y}'_{p})^{N-p} \\ \times \prod_{i=1}^{p}\prod_{j=i+1}^{p}\mathrm{sgn}(y_{i}-y_{j})\mathrm{sgn}(y'_{i}-y'_{j}), 
 \end{multline}
where
\begin{align}
\mathrm{P}_0(\boldsymbol{y}_{p},\boldsymbol{y}'_{p})=\int_0^1\mathrm{d}\xi\,\prod_{i=1}^{p}\mathrm{sgn}(y_{i}-\xi)\mathrm{sgn}(y'_{i}-\xi). \nonumber
\end{align}
The transformation between eigenorbitals is also analogous, $u_k(\boldsymbol{x}_p)=\left[\Pi_{i=1}^p\phi(x_i)\right]v_k(F(\boldsymbol{x}_p))$.
The rigorous argumentation presented above shows that eigenvalues of any $p$-RDM calculated for strongly attractive $\mathfrak{p}$-wave fermions have universal values and are completely independent of the shape of an external potential. 

\begin{figure}
\includegraphics[width=\linewidth]{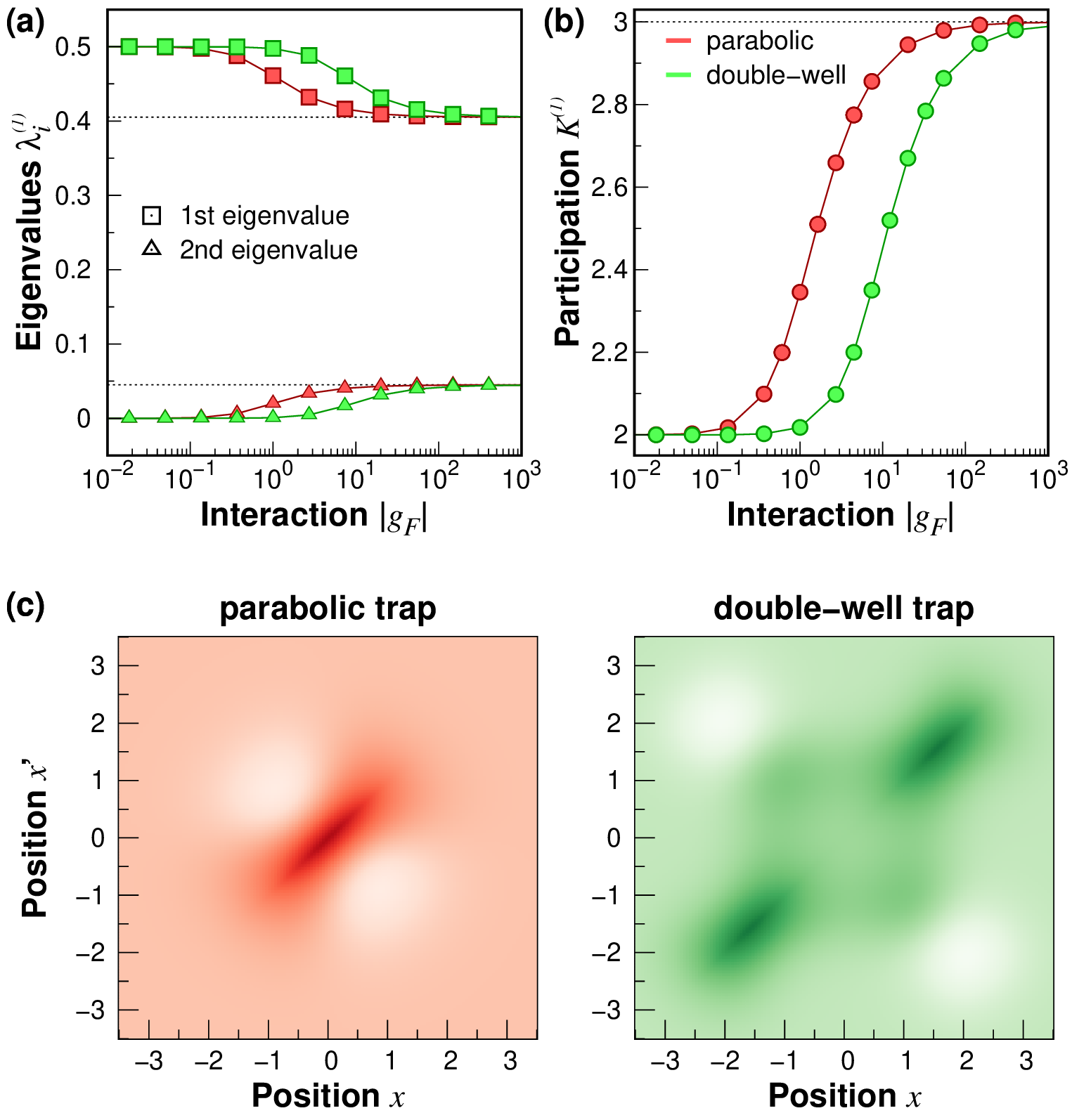}
\caption{Spectral properties of $1$-RDM for the system of $N=2$ $\mathfrak{p}$-wave fermions confined in a parabolic trap (red) and double-well potential (green). {\bf (a)} The two largest eigenvalues (doubly degenerated) as functions of interactions. For finite interactions, spectral decomposition depends on the shape of external trapping. However, in the limit of infinite attractions, all eigenvalues saturate on corresponding universal values (dashed lines, here $4/\pi^2$ and $4/9\pi^2$). {\bf (b)} Participation number $K^{(1)}$ as a function of interaction. Depending on external potential, the number grows monotonically with different slopes. However, independently of confinement, it always saturates at the universal value (dashed line at $K^{(1)}_{\infty}=3$). {\bf (c)} $1$-RDM in the position domain $\rho^{(1)}(x,x')$ obtained in the limit of infinite attractions $g_F\rightarrow -\infty$ for two different confinements. Although these two density matrices are substantially different, due to the universality proven, they have exactly the same eigenvalues.
\label{Fig1}}
\end{figure}

{\bf Minimalistic example.} The proven universality has not only fundamental meaning but brings also practical consequences since it provides the universal upper bounds on multi-component correlations forced by interactions in any one-dimensional $\mathfrak{p}$-wave fermionic system. To illustrate that let us consider the simplest system of $N=2$ fermions confined in two, very different potentials, {\it i.e.}, pure parabolic trap, $V(x)=x^2/2$, and a deep double-well trap, $V(x)=x^2/2 +4\mathrm{e}^{-x^2/2}$. In these cases, the ground-state wave function can be easily obtained numerically for any interaction strength $g_F$ (via mapping from two-boson solutions). Simultaneously, all correlations between particles are encoded in corresponding $1$-RDM which can be quite easily diagonalized numerically on a dense grid for any interaction strength $g_F$~\cite{2020KoscikNJP}.

Of course, in the non-interacting case, the ground state wave function is provided by a single Slater determinant of the two lowest single-particle orbitals. Thus, the $1$-RDM has only one non-zero doubly degenerated eigenvalue equal to $1/2$ (Fig.~\ref{Fig1}a). Then, along with increasing attractive interactions, other single particle orbitals start to contribute to the ground state, their eigenvalues become non-zero, the participation number grows monotonically (Fig.~\ref{Fig1}b), and the state becomes entangled~\cite{2001LiPRA,2006SunPRA}. It is clear that for a given interaction strength the number $K^{(1)}$ depends also on the shape of an external potential (red and green curves for parabolic and double-well potentials, respectively). In the limit of infinite attractions ($g_F\rightarrow -\infty$), however, corresponding eigenvalues of $1$-RDM become exactly the same and the participation $K^{(1)}$ saturates on the universal value. Concurrently, corresponding $1$-RDMs are completely different and have significantly distinct spatial shapes (see Fig.~\ref{Fig1}c). It turns out that in this case, all the spectral properties of the $1$-RDM  can be found analytically, {\it i.e.}, one can find exact solutions of the eigenproblem \eqref{newep}. All the eigenvalues of $\rho_0^{(1)}(y,y')$ are doubly degenerated and equal to $\lambda^{(1)}_k = [2/\pi(2k-1)]^2$, while corresponding eigenorbitals have a form $v_k^{(+)}(y)=\sqrt{2}\sin[(2k-1)\pi y]$ and $v_k^{(-)}(y)=\sqrt{2}\cos[(2k-1)\pi y]$. It means that the dominant eigenvalue $\lambda_1^{(1)}$ and the participation $K^{(1)}$ saturate at $4/\pi^2$ and $3$, respectively (horizontal dashed lines in Fig.~\ref{Fig1}a and Fig.~\ref{Fig1}b). They agree with previous results obtained for box potential~\cite{2007HaoPRA}. Let us mention here that in the case of two bosons confined in these potentials, $1$-RDM is no longer universal. In the limit of infinitely strong repulsions, the two largest eigenvalues are (0.7735, 0.1712) and (0.5847, 0.4127) (participations 1.583 and 1.952) for the harmonic and the double-well potential, respectively.

\begin{figure}
\includegraphics{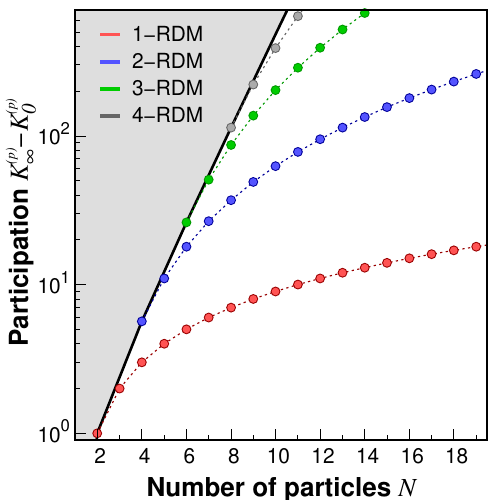}
\caption{Universal participation number $K_\infty^{(p)}$ calculated for four the lowest $p$-RDMs and different numbers of particles $N$ (different colors). The solid black line corresponds to the participation obtained for the bipartition of the system ($p=N/2$). For convenience, in all the cases, we display the difference between $K_\infty^{(p)}$ and its value for a non-interacting system, $K_0^{(p)}=N!/p!(N-p)!$.\label{Fig2}}
\end{figure}

{\bf High-order correlations.} The proven spectral universality of reduced density matrices maybe also utilized when higher-order correlations (encoded in higher $p$-RDMs) are considered. For some purposes, it does not even require solving the universal eigenproblem \eqref{eigenprob2}. For example, thanks to \eqref{rhopp}, one can straightforwardly show that the purity ${\cal P}^{(p)}$ may be expressed as a pure polynomial multiple integral of the form 
\begin{multline}\label{app}
{\cal P}^{(p)}=\int_{\boldsymbol{0}}^{\boldsymbol{1}}\!\mathrm{d}\boldsymbol{y}_p\int_{\boldsymbol{0}}^{\boldsymbol{1}}\!\mathrm{d}\boldsymbol{y}_p'\,\mathrm{P}_0(\boldsymbol{y}_{p},\boldsymbol{y}'_{p})^{2(N-p)}\\
=(2p)!\int_{0}^{1}\!\!\mathrm{d}y_{1}\!\int_{y_1}^{1}\!\!\mathrm{d}y'_{1}\!\int_{y_1'}^{1}\!\!\mathrm{d}y_{2}\!\int_{y_2}^{1}\!\!\mathrm{d}y'_{2}\ldots\\ \ldots\int_{y'_{p-1}}^{1}\!\!\!\!\!\!\mathrm{d}y_{p}\!\int_{y_p}^{1}\!\!\mathrm{d}y'_{p}
 \left[1+2\sum_{i=1}^p(y_{i}-y_{i}^{'})\right]^{2(N-p)}.
\end{multline}
In principle, it means that in the limit of infinite attractions, the purity can be calculated analytically for any $N$ and $p$ (of course independently of the shape of an external potential). In practice, calculating this kind of integrals is arduous and can be performed only with symbolic calculus software. As an instructive example, in Fig.~\ref{Fig2} we show the participation $K^{(p)}_\infty$ calculated analytically in this limit for $p=1,\ldots,4$ (different colors) and $p=N/2$ (solid black line) as functions of the number of particles $N$. The latter gives the upper limit for all lower-$p$ participations. In this way we determine rigorously, the amount of correlation that can be obtained in systems of strongly interacting $\mathfrak{p}$-wave fermions regardless of the shape of the trapping potential. It is evident that the number of eigenorbitals contributing significantly in the decomposition \eqref{diag} grows rapidly with $N$ and $p$ signaling a quick boost of quantum correlations.  We find that depending on $p$, the participation changes from linear ($K_\infty^{(1)}=2N-1$) to exponential-like (numerical fit provides approximate dependence $K_\infty^{(N/2)}\sim 2^{N}$).  

{\bf Final remarks.}
In principle, the identified universality holds for any number of particles. However, since the gap to excited many-body states decreases with growing $N$, for sufficiently large particle numbers experimental observation of the universality may be challenging. This obviously depends also on the shape of confinement.

Universal features of quantum correlations are direct consequences of the form of the ground state~\eqref{wf}. Therefore, the reasoning can be easily applied to any state having such a representation, using an appropriate cumulative distribution. Unfortunately, excited states of the system under study do not necessarily have such a representation. Therefore, the question of whether their correlations manifest any universality remains open and requires further investigation.

Finally, let us also mention that our results suggest that the universality can be used as a tool to indicate states that cannot be achieved as a many-body ground state. For example, as noticed previously in a two-particle system, the participation $K^{(1)}$ never exceeds universal value $K_\infty^{(1)}=3$. If this is true, one can immediately argue that states with larger $K^{(1)}$, potentially interesting from the quantum information point of view, are not achievable as the ground state of $N=2$ $\mathfrak{p}$-wave fermions independently on interaction strength and shape of the confinement. One of such states is the entangled Slater-rank-two state~\cite{2001LiPRA,2001SchliemannPRA,2006SunPRA} build as an equal superposition of two distinct Slater determinants (the participation $K^{(1)}$ for this state is equal $4$). Since the universal behavior is formulated for any $p$-RDM, similar reasoning can be also applied for higher-order correlations when a larger number of particles is considered. In fact, for a given number of particles $N$, considering all possible upper bounds derived for different $p$ significantly reduces the space of many-body states attainable in the interacting ground state of $\mathfrak{p}$-wave fermions. From this point of view, deeper studying of all limitations forced by recognized universality may bring a better understanding of different $\mathfrak{p}$-wave fermions systems also for finite interactions. 
 
{\bf Acknowledgements.} This work by TS was supported as part of a project funded by the Polish Ministry of Education and Science on the basis of agreement no. 2022/WK/06.

\bibliography{biblio}

\end{document}